\documentclass[preprint,twoside,a4paper,11pt]{article}

\usepackage{blindtext}

\usepackage{irrj2e}

\renewcommand{\cite}{\citep}

\usepackage{caption}
\usepackage{adjustbox} %
\usepackage{tikz,pgfplots}
\usepackage{arydshln}
\usepackage{stfloats}
\usepackage{booktabs}
\usepackage{amsmath}
\usepackage{float}

\usepackage{lastpage}

\irrjheading{1}{2025}{1-\pageref{LastPage}}{2/25}{-}{10.54195/irrj.00000}

\ShortHeadings{The Role of Vocabularies in Learning Sparse Representations for Ranking}{Kim, Lee and Won}
\firstpageno{1}

\begin{document}

\author{\name Hiun Kim \email hiun.kim@navercorp.com \\
       \addr Naver\\
       Seongnam, South Korea
       \AND
        \name Tae Kwan Lee \email taekwan.lee@navercorp.com \\
       \addr Naver\\
       Seongnam, South Korea
       \AND
        \name Taeryun Won \email lory.tail@navercorp.com \\
       \addr Naver\\
       Seongnam, South Korea}

\editor{My editor}

\title{The Role of Vocabularies in Learning Sparse Representations for Ranking}

\maketitle

\begin{abstract}
Learned Sparse Retrieval (LSR) such as SPLADE has growing interest for effective semantic 1st stage matching while enjoying the efficiency of inverted indices.
A recent work on learning SPLADE models with expanded vocabularies (ESPLADE) was proposed to represent queries and documents into a sparse space of custom vocabulary which have different levels of vocabularic granularity.
Within this effort, however, there have not been many studies on the role of vocabulary in SPLADE models and their relationship to retrieval efficiency and effectiveness.
~

To study this, we construct BERT models with 100K-sized output vocabularies, one initialized with the ESPLADE pretraining method and one initialized randomly.
After finetune on real-world search click logs, we applied logit score-based queries and documents pruning to max size for further balancing efficiency.
The experimental result in our evaluation set shows that, when pruning is applied, the two models are effective compared to the 32K-sized normal SPLADE model in the computational budget under the BM25.
And the ESPLADE models are more effective than the random vocab model, while having a similar retrieval cost. 
~

The result indicates that the size and pretrained weight of output vocabularies play the role of configuring the representational specification for queries, documents, and their interactions in the retrieval engine, beyond their original meaning and purposes in NLP.
These findings can provide a new room for improvement for LSR by identifying the importance of representational specification from vocabulary configuration for efficient and effective retrieval.
\end{abstract}

\begin{keywords}
neural networks, sparse representations, vocabularies, ranking
\end{keywords}

\section{INTRODUCTION}

Neural matching tries to overcome the vocabulary mismatch problem by projecting the semantics of queries and documents into an embedding space for matching.
Many approaches for neural matching have been proposed with different modeling precision and volumes of target documents.

\cite{nogueira2019passage} proposed a cross-encoder approach that models interactions of queries~(Q) and documents~(D) representations within the latent space of BERT \cite{devlin2019bert}. This approach requires calculating Q-D relevancy scores for all Q-D pairs obtained by 1st matching, hence it works as a reranker.
The computational burden of the cross-encoder approach hinders applying neural matching for 1st matching.

The bi-encoder approach helps to resolve the scalability issue by representing queries and documents in the same embedding space.
This allows encoding the D offline and the Q online.
For example, \cite{reimers2019sentence} proposed Sentence-BERT, a sentence-level bi-encoder, and \cite{khattab2020colbert} proposed ColBERT, a token-level bi-encoder.
The bi-encoder approach shows wide adoption and has been extensively studied \cite{xiong2020approximate, lin2021batch, hofstatter2020improving, hofstatter2021efficiently, qu2020rocketqa, karpukhin2020dense}; its serving efficiency alleviates its decreased model effectiveness from the embedding space bottleneck, as Q and D are encoded separately and later interact in their unified embedding space.

However, one of the central challenges of the existing bi-encoder approach is still the computational cost at retrieval.
Most bi-encoders make pooled "dense" representations of Q, D.
This requires vector retrieval systems like Faiss \cite{johnson2019billion} for retrieval, which is expensive compared to inverted index systems.
Also, such dense retrieval systems are difficult to integrate with existing inverted index systems in the 1st matching stage, as retrieval is performed separately, hindering both high-recall 1st stage matching and combining relevance signals from dense vector and other features in the inverted index at the 1st stage matching.

Within this context, Learned Sparse Retrieval (LSR) \cite{nguyen2023unified, dai2019context, nogueira2019document, bai2020sparterm, gao2021coil, mallia2021learning} is an approach to perform bi-encoder based neural matching in homogeneous inverted index systems, towards efficient neural matching and ease of signal engineering at the ranker.
\cite{formal2021splade} proposed SPLADE, a bi-encoder approach for neural matching using learned sparse representations.
Similar to the dense retrieval approach, SPLADE uses the same discriminative language model, BERT.
SPLADE initiates its training by obtaining the discriminative representations of Q, D by pooling its Masked Language Modeling (MLM) outputs that are originated from contextualized word embeddings of BERT, with metric learning and sparsity regularization loss.

MLM is a self-supervisory task for learning the context of sentences by filling the blank using cross-entropy classification objectives \cite{devlin2019bert}.
Many follow-up studies on SPLADE \cite{formal2022distillation, lassance2022efficiency, lassance2023static, nguyen2023unified, dudek2023learning, qiao2023representation, kong2023sparseembed, formal2024splate} show that the pretrained weights of BERT and the use of MLM head outputs provide suitable foundations for finetuning using ranking loss with sparsity regularization.

However, industrial deployment of SPLADE is still costly compared to the traditional retrieval method, BM25 \cite{robertson2009probabilistic}.
One primary reason for the pitfall of plain SPLADE models is that,
compared to the natural language terms used in the BM25, the sparsity of representations, hence the diversity of activated terms, can be low. (i.e., have larger co-occurrence Wordpiece terms in the corpus, compared to the natural language terms).
To some degree, this seems inevitable; the learned, hence non-linearly transformed representations need to have higher commonalities with others to achieve high-recall retrieval compared to the literal lexical representations.
For example, Wordpiece vocabularies usually have more coarse-grained meaning (e.g., app, \#\#le) compared to natural language terms (e.g., apple)\footnote{The count difference between Wordpiece word vocab and natural language word vocab can exhibit this quantitatively.}.
The coarse vocabularic granularity of plain SPLADE models introduces higher term co-occurrence between data, which can provide effectiveness for retrieval, for instance, by allowing to alleviate OOV issues, but inevitably leads to a "small vocabulary set-long posting list" situation of inverted index, which influences the higher computational cost of retrieval.

One of the recent works on studying vocabularies in SPLADE models is Expanded-SPLADE from \cite{dudek2023learning}, which presents a pretraining approach to learning custom vocabularies for retrieval. 
Within this effort, however, there have not been many studies on the role of vocabulary in SPLADE models and their relationship to retrieval efficiency and effectiveness.
~

To study this, we construct BERT models with 100K-sized output vocabularies, one initialized with the ESPLADE pretraining method and one initialized randomly.
After finetune on real-world search click logs, we applied logit score-based static term pruning \cite{lassance2023static} for both queries and documents pruning to max size for further balancing efficiency.
The experimental result in our evaluation set shows that, when pruning is applied, the two models are effective compared to the 32K-sized normal SPLADE model in the computational budget under the BM25.
And the ESPLADE models are more effective than the random vocab model, while having a similar retrieval cost. 
~

The result indicates that the size and pretrained weight of output vocabularies play the role of configuring the representational specification for Q, D, and their interactions in the retrieval engine, beyond their original meaning and purposes in NLP.
These findings can provide a new room for improvement for LSR by identifying the importance of representational specification from vocabulary configuration for efficient and effective retrieval.

\section{RELATED WORK}

\textbf{Vocabularies and the SPLADE Finetuning Performance}
\cite{mackenzie2023exploring} assess the representation powers of SPLADE models with different controlled vocabularies.
Specifically, this paper discusses that the size of vocabulary and the pretrained weight of MLM heads are important factors in the finetuning performance of SPLADE models.
However, this study is conducted using a limited vocabulary, which includes only part of vocab from the existing vocabulary.
Hence, the study has not yet discussed the expansion mechanism of vocabularies and performance trend beyond the size of the existing vocabulary, and ways to reveal such performance.

\textbf{Learning Custom Vocabularies in SPLADE Models}
ESPLADE \cite{dudek2023learning} shows an approach to expand vocabulary for an arbitrary set of vocabulary using expanded masked language modeling training.
This approach leverages the pretrained weights of Wordpiece-based plain SPLADE models to construct new vocabulary with different semantics and dimensions.
While the study shows the ESPLADE models maintain effectiveness and efficiency compared to the original SPLADE models, ways to leverage such expanded vocabulary for efficiency have not been studied.

\cite{yu2024improved} present an approach to improve the effectiveness and efficiency of LSR models by constructing Wordpiece vocabulary from the target corpus and pretraining the BERT from scratch on the target corpus.
This study enables training a larger dimension of vocabularies, but the effect of static pruning on SPLADE Models with different vocabulary sizes remains unstudied.

\textbf{Sparsification and Efficiency of SPLADE Models}
SPLADE \cite{formal2021splade} models employ the FLOPS regularization loss \cite{paria2020minimizing} to sparsify query and document vectors while minimizing the ranking loss. The ESPLADE models \cite{dudek2023learning} propose a joint FLOPS regularization loss, which sparsifies the intersection of query and document, which are more directly related to the retrieval efficiency.
\cite{yang2021sparsifying} present ways to learning more sparsified representations by finetuning with modified pooled BERT logits that only use the top-k highest scores along with original pooled BERT logits.
In the inference phase, \cite{lassance2023static} present that applying the concepts of term pruning techniques \cite{buttcher2006document, thota2011within} to LSR is helpful for efficient retrieval with marginal sacrifice on the effectiveness.

\section{ESPLADE TRAINING}

This section describes pretraining, finetuning, and evaluation of the ESPLADE models.
Throughout this paper, we refer pretrained model as EMLM (Expanded Masked Language Model) and the finetuned model as ESPLADE.
Both model was introduced by \citet{dudek2023learning}, and our procedure is based on that paper.

Here, we introduce their method at a high level.
The prediction label of the MLM task is a token of vocabulary.
The MLM outputs the probability of each token of the given vocabulary.
The specification of vocabulary is shared physically in the model by the number of word embedding vectors in the input lookup table corresponding to each word in the vocabulary, and the weight vector of the output MLM FC layer for the prediction, also corresponding to each word in the vocabulary.
Note that weight tying~\citep{press2017using} is often applied for sharing weight between them.

In the \citet{dudek2023learning}'s method, EMLM reconstructs the weight and bias of the output MLM FC layer, leveraging the original MLM FC layer. The new output vocabulary can have a different set of words, and a different total number of words~(or dimension).

\subsection{Pretraining}

\subsubsection{Model Construction and Initialization}

We used an in-house BERT model that has 6 layers, 768 hidden dimensions, 12 heads, 55M parameters, and a vocab size of 32001. The model is pretrained from corpora that include news, books, wikis, and encyclopedias centric primarily to Korean. The BERT model uses a subword tokenizer. We set the maximum length accepted by the model to 64.

We refer to the procedure on \cite{dudek2023learning} to construct and initialize EMLM models.
Here we describe the procedure for illustrative purposes.
First, given the corpus, we split the corpus into a train and a validation set that are exclusive.
Then get the top 100K frequentist vocabulary after extracting all of the unigrams in the corpus and sorted by their frequency.
Note that each vocabulary must be tokenized into one or more Wordpiece subwords.
This vocabulary is the expanded vocabulary set $U$.
We construct the weight and bias of the EMLM head by performing subword tokenization on each vocabulary word $U$.
For instance, the term "love" can be compounded by two Wordpiece subword vocabulary "lo" and "\#\#ve", then we mean pooling the two words' weight and bias of MLM FC head to construct the weight and bias of expanded vocabulary "love" in the $U$.
We replace the new MLM head layer with a constructed weight and bias of 100K $U$ terms.

\subsubsection{Pretraining Masking Strategy}

We mask $U$ vocabulary terms by the ratio of 15\% in each title.
For instance, a sentence that has 20 $U$ terms, we mask and only mask Wordpiece terms corresponding to the selected 3 $U$ terms.
Note that this masking strategy is different from conducting the Bernoulli trial for each $U$ term by 15\%.
We guarantee at least 1 $U$ term to be masked if the count of 15\% $U$ terms is below 1 (i.e., short titles).
We follow the masking strategies details of the original BERT paper \cite{devlin2019bert}, 80\% of target mask tokens will be transformed into \texttt{[MASK]} token, 10\% random token, 10\% unchanged.
If the target mask $U$ token consists of multiple Wordpiece tokens, we mask all subwords (i.e., mask the first and all later-positioned subwords).
The target mask token will have the label of the corresponding $U$ token.

\subsubsection{Hyperparameters and Trained Dataset Volume}

We used the learning rate of 1.6e-4 with linear learning rate scheduling.
We used the warmup of 10,000 steps.
The batch size is 1024 with 8 A100 GPUs.
We used AdamW optimizer with weight decay of 0.01.
We used 16-bit (mixed) precision training.

An EMLM model for this study was trained to 2.9M steps, and we use the 600K steps' checkpoint.
During training of 600K steps, the model sees a total of 614 million records, which are all unique web titles from our in-house collection.
For the expanded vocabulary of the EMLM model, we have constructed the 100K most frequent unigrams from the finetuning trainset. This is to align the vocabulary's distribution to the trainset while pretraining on a larger corpus.

\begin{table*}
    \centering

\footnotesize
\begin{tabular}{lrrrrr}
\toprule
dataset & \# of q-title pairs & \# of unique titles & \# of queries & docs per query & queries per doc\\
\midrule
trainset & 262,959,262 & 114,139,846 & 34,145,679 & 7.701 & 2.304\\
testset & 9,629,311 & 8,210,573 & 1,198,184 & 8.036 & 1.070\\
validset & 339,851 & 307,901 & 39,998 & 8.496 & 1.004\\
\bottomrule
\end{tabular}
\caption{\textbf{Stats for Datasets.} \small{validset and testset include distractor queries and documents.}}
\label{tab:stats-set}

\end{table*}

\subsection{Finetuning Datasets}

The model was trained on a dataset of queries and web document titles.
We used part of our search log that contains users' query-document (Q-D) clicks\footnote{Due to the diverse nature of queries and documents that we possess, the dataset is multilingual, but centric primarily to Korean. There have been concepts in quantitative linguistics, such as Zipf's law, and recent work \cite{zoph2016transfer, johnson2017google, feng2020language} in the NLP community identifies the similarity in the latent structure of different languages, by emphasizing the transferability between different languages in NLP tasks. Although further experimentation is fruitful, we foresee the findings of this paper can be applicable to other languages as well.}.

Table \ref{tab:stats-set} shows stats for the dataset.
To obtain a clear supervision signal, we use only when there are multiple documents per query that appear within the upper part of the search results and have been clicked on multiple times. 
We allow the same document title to appear multiple times in the dataset.
For constructing the testset and validset, we use our in-house SBERT-based sentence embedding model to cluster these datasets into 10K clusters based on the semantics of queries.

\subsubsection{Testset}
Given the clustered result, we randomly pick 30 queries from each cluster (a total of 300K queries) and use these queries and associated documents as labels.
For the distractor, we randomly pick  90 queries from each cluster and get their documents. (documents from a total of 900K queries).
We exclude documents that have already been picked as positive for a distractor.

\subsubsection{Validset}
Given the clustered result, we randomly pick 1 query from each cluster (a total of 10K queries) and use these queries and associated documents as labels.
For the distractor, we randomly pick 3 queries from each cluster and get their documents. (documents from a total of 30K queries).
We exclude documents that have already been picked as positive for a distractor.

\subsubsection{Trainset}
Given the clustered result, we pick all queries from each cluster to obtain the trainset-small (stats of this set are described in the Table \ref{tab:stats-set2} at the end of the main text).
We construct the trainset by adding additional search logs into trainset-small.
For both sets, we exclude queries that are used in the positive and distractor of the testset and the validset.
Titles appeared in the positive and distractor of both testset, and the validset can be included.

\subsection{Finetuning Details}

\subsubsection{Losses}

We finetuned the EMLM model with a finetuning dataset for the ranking objective to obtain an ESPLADE model.

Unlike the original SPLADE paper, models were trained using only in-batch negative loss, which is formulated as:
\begin{equation}
\label{eq:1}
    \mathcal{L}_{\text{in-batch}} = - \frac{1}{|B|} \sum_{(q,d^+)} \log \frac{e^{s(q,d^+)}}{e^{s(q,d^+)} + \sum\limits_{d^- \in B} e^{s(q,d^-)}}
\end{equation}
where $s(q,d)$ represents the similarity score between query $q$ and document $d$, $d^+$ is the relevant document, $d^-$ are negative samples drawn from the batch $B$, and $|B|$ is the batch size. This decision was made to simplify the sampling process.
As a result, the Equation \ref{eq:1} is equivalent to the N-pair loss objective \cite{sohn2016improved}.

For regularization, the SPLADE model employs the FLOPS loss, which is defined as:
\begin{equation}
    \mathcal{L}_{\text{FLOPS}}(T) = \bar{w}^{(T)} \cdot \bar{w}^{(T)}
\end{equation}
where $\bar{w}^{(T)}$ denotes the mean representation of the output embeddings across all texts in $T$, and $\cdot$ is the inner product \cite{paria2020minimizing}. This loss applies separately to each of the Q, D representation batches.

In contrast, the ESPLADE model utilizes the joint FLOPS regularization loss, which is defined as:
\begin{equation}
    \mathcal{L}_{\text{jFLOPS}}(Q, D) = \bar{w}^{(Q)} \cdot \bar{w}^{(D)}
\end{equation}

This formulation ensures that the embeddings of queries and documents are effectively regularized, aligning their representations in the embedding space for improved retrieval efficiency.

\subsubsection{Hyperparameters}

We used the learning rate of 1e-4 with a linear learning rate scheduler. We use the warmup steps of 10,000.
For the optimizer, we used AdamW with weight decay of 0.01.
We used 16-bit (mixed) precision training.
For models with 100K-sized vocab (esplade, rand), we trained with a 3072 batch size to 400K steps.
For the plain SPLADE model, which has a 32K-sized vocab dimension, we trained with a 7168 batch size to 200K steps.

Note that in the plain SPLADE model, we use the lower steps compared to the 100K-sized vocab models, because in our experiment, trials of 400K steps cause overfitting.
Also, the difference in batch size per vocab size was caused by the physical limit of our GPU memory, as a larger vocab size introduces a larger MLM logit vector per individual input Wordpiece tokens \footnote{In N-pair loss objective, we observe that the smaller batch size negatively impacts the model effectiveness (see Table \ref{tab:finetuning-result} at the end of the main text). Thus, we expect the model's effectiveness solely from the loss is lower for models with a 100K-sized vocab compared to models with a 32K-sized vocab.}. The validset and testset results of finetuned models are presented in Table \ref{tab:finetuning-result} at the end of the main text.

\subsection{Q, D Term Pruning with MLM Logit Score}

The ranking loss and FLOPS regularization loss can be opposing forces in the training, each aiming to learn effective and efficient representations, respectively.
Since the two losses are combined in the finetuning, it can be difficult to obtain representations of maximal efficiency alone, as this tends to increase the ranking loss by decreasing the amount of information in the representations.

We applied static pruning methods for LSR \cite{lassance2023static} to obtain further retrieval efficiency. Specifically, we remove low scored terms based on a max size from the learned sparse representations pooled from the MLM output of BERT models, where the score is the finetuned MLM logit score.
In previous work \cite{buttcher2006document}, the target of static pruning is to remove terms in the inverted index, and the static pruning for LSR follows the same \cite{lassance2023static}.

However, different from the traditional lexical sparse retrieval, the terms of LSR are learned, and the bi-encoder structure is trained to project Q and D into the same embedding space. Hence, the gap between the Q and D term counts tends to be closer, compared to the lexical Q and D term counts, as Figure \ref{tab:201} and Table \ref{tab:eval-set} show respectively.
Due to this, we applied static pruning for both documents as well as queries based on max size to further obtain retrieval efficiency, which is similar to the first matching setup discussed in \cite{lassance2024two}.
Throughout this paper, we use qk and dk, which show the max size of Q and D terms to represent each Q and D, which only includes the terms with the top-k highest MLM logit score.

\section{EVALUATION}

\begin{table*}
    \centering

\footnotesize
  \begin{tabular}{c|r}
    \toprule
    Document Count & 20,372,952 \\
    \midrule
    Test Query Count & 8,936 \\
    \midrule
    Avg. Document Title Length & 13.4 words \\
    \midrule
    Avg. Query Length & 2.6 words \\
  \bottomrule
\end{tabular}
\caption{\textbf{Stats for Evaluation Set.}}
  \label{tab:eval-set}

\end{table*}

\subsection{Evaluation Set}

We constructed an additional larger scale evaluation set to better capture the semantic representation power of models.
Table \ref{tab:eval-set} shows the stats for the evaluation set.

\subsubsection{Sampling Queries}
We sampled triplets of query, positive document list, and negative document list from the log.
To select the evaluation query set $Q$, we cluster queries using an embedding model.
We use this clustering result to pick candidate queries that are more semantically balanced.
We sample these candidate queries to construct the $Q$.
We picking up queries based on the matched term ratios to their corresponding positive documents (titles). Note that among matched term ratios of multiple positive documents, we use the highest.
Based on that value, we classify queries into three quantiles - low, middle, and high matched term ratios.
We pick the same number of queries in each quantile.
This was performed to obtain queries that have a diverse level of lexical matching difficulty.

\subsubsection{Sampling Positive and Negative Documents for Query}

For positives, we select the top-k positive documents based on label score. For each query, we pick 3 documents at most.
For negatives, we select negative documents for each target query from positive documents of queries in other clusters that above a fixed value of high term matching ratio to the target query.
(i.e., naive lexical matching hurts the performance).
For each target query, we obtain negative documents by selecting the top 10 documents at most that have the highest term matching ratio to the target query.
We only use triplets where all query, positive documents, and negative documents exist.

\subsection{Retrieval Efficiency Measure}
To evaluate retrieval efficiency, we define the FLOPS metric that approximates the volume of 1st stage matched documents in the search engine by combining of posting list lengths for each query in the evaluation query set.

This FLOPS metric measures the computational cost of retrieval on a search engine, as it approximates the traversal cost of posting lists, which reflects the traversal cost itself, and the cost of more expensive later-stage ranking proportional to the volume of 1st matched documents. The FLOPS metric is defined as:

\begin{equation}
\text{FLOPS} = \frac{\sum_{q \in Q} \sum_{t \in q} \left| P_t \right|}{|Q| \cdot |D|}
\end{equation}

Where $Q$ is the set of evaluation queries, $t$ is a query term in the individual evaluation query.
$|P_t|$ is the posting list length, which is the number of matching documents belonging to term $t$.
$|Q|$ and $|D|$ are the total count of queries and documents which used for the normalization.

\begin{figure*}
\begin{minipage}{0.55\textwidth} %
    \centering
    \footnotesize
    \setlength{\tabcolsep}{2.5pt}

  \begin{tabular}{lccccccc}
    \toprule

model & qk & dk & L0\_q & L0\_d & FLOPS & MRR@10 & R@10 \\
\midrule
bm25 & . & . & . & . & 0.00277 & 0.203 & 0.2527 \\
\midrule
\multicolumn{8}{l}{Q, D Not Pruned} \\
splade-32K & 0 & 0 & 11.35 & 31.17 & 0.02173 & 0.2733 & 0.3543 \\
rand-100K & 0 & 0 & 12.16 & 34.27 & 0.0128 & 0.2484 & 0.3299 \\
esplade-100K & 0 & 0 & 9.77 & 39.9 & 0.01 & 0.2549 & 0.3347 \\
\midrule
\multicolumn{8}{l}{Q, D Pruned (FLOPS near 0.003x)} \\
splade-32K & 5 & 10 & 4.95 & 9.93 & 0.00383 & 0.2643 & 0.3281 \\
rand-100K  & 5 & 20 & 4.96 & 18.99 & 0.0039 & 0.2395 & 0.3128 \\
esplade-100K & 7 & 20 & 6.31 & 18.29 & 0.0047 & 0.2579 & 0.3359 \\
esplade-100K & 5 & 20 & 4.84 & 18.29 & 0.0037 & 0.2491 & 0.327 \\
\midrule
\multicolumn{8}{l}{Q, D Pruned (FLOPS under bm25; near 0.002x)} \\
splade-32K & 5 & 5 & 4.95 & 5 & 0.00194 & 0.2243 & 0.2621 \\
rand-100K & 5 & 10 & 4.96 & 9.96 & 0.0022 & 0.2371 & 0.2981 \\
esplade-100K & 5 & 10 & 4.84 & 9.91 & 0.0022 & 0.251 & 0.3189 \\

  \bottomrule
\end{tabular}

\end{minipage}%
\begin{minipage}{0.03\textwidth}
~
\end{minipage}
\begin{minipage}{0.44\textwidth} %
    \centering

        \begin{tikzpicture}
            \begin{axis}[
                xlabel={FLOPS},
                ylabel={MRR@10}, %
                xlabel near ticks,
                ylabel near ticks,
                grid=major,
                width=7.5cm, %
                height=4cm,%
                legend pos=south east,
                minor y tick num=2,  %
                minor grid style={lightgray,very thin},
                grid=both,
                label style={font=\footnotesize},
                tick label style={font=\footnotesize},
                ylabel style={at={(0,1)}, anchor=south,rotate=-90},
                legend style={
                    inner sep=1pt,
                    nodes={inner sep=1pt},
                    xshift=5pt,
                    yshift=-2pt,
                    font=\scriptsize,
                },
            ]
                \addplot[color=blue, mark=*, thick] coordinates {
                    (0.02173, 0.2733)
                    (0.00383, 0.2643)
                    (0.00194, 0.2243)
                };
                \addlegendentry{splade-32K}
                
                \addplot[color=orange, mark=triangle*, thick] coordinates {
                    (0.0128, 0.2484)
                    (0.0039, 0.2395)
                    (0.0022, 0.2371)
                };
                
                \addlegendentry{rand-100K}
                \addplot[color=red, mark=square*, thick] coordinates {
                    (0.01, 0.2549)
                    (0.0037, 0.2491)
                    (0.0022, 0.251)
                };
                \addlegendentry{esplade-100K}
            \end{axis}
        \end{tikzpicture}

        \begin{tikzpicture}
            \begin{axis}[
                xlabel={FLOPS},
                ylabel={R@10}, %
                xlabel near ticks,
                ylabel near ticks,
                grid=major,
                width=7.5cm, %
                height=4cm,%
                legend pos=south east,
                minor y tick num=2,  %
                minor grid style={lightgray,very thin},
                grid=both,
                label style={font=\footnotesize},
                tick label style={font=\footnotesize},
                ylabel style={at={(0,1)}, anchor=south,rotate=-90},
                legend style={
                    inner sep=1pt,
                    nodes={inner sep=1pt},
                    xshift=5pt,
                    yshift=-2pt,
                    font=\scriptsize,
                },
            ]
                \addplot[color=blue, mark=*, thick] coordinates {
                    (0.02173, 0.3543)
                    (0.00383, 0.3281)
                    (0.00194, 0.2621)
                };
                \addlegendentry{splade-32K}
                
                \addplot[color=orange, mark=triangle*, thick] coordinates {
                    (0.0128, 0.3299)
                    (0.0039, 0.3128)
                    (0.0022, 0.2981)
                };
                \addlegendentry{rand-100K}
                
                \addplot[color=red, mark=square*, thick] coordinates {
                    (0.01, 0.3347)
                    (0.0037, 0.327)
                    (0.0022, 0.3189)
                };
                \addlegendentry{esplade-100K}
            \end{axis}
        \end{tikzpicture}

\end{minipage}
\caption{\textbf{Result on Evaluation Set.} \small{Models are trained on trainset. For the splade-32K model (32K output vocab), we trained with top-k masking for q\_K=500, d\_K=1000. For other models (100K output vocab), we trained with top-k masking for q\_K=1000, d\_K=2000. The qk and dk are the max size of Q and D terms to represent each Q and D, where the terms with top-k highest MLM logit score are included (value of 0 means unpruned). L0\_q, L0\_d denote the average number of Q, D terms used for the evaluation after the pruning is applied. The right figure shows the visualization of the left table.}}
  \label{tab:201}
\end{figure*}
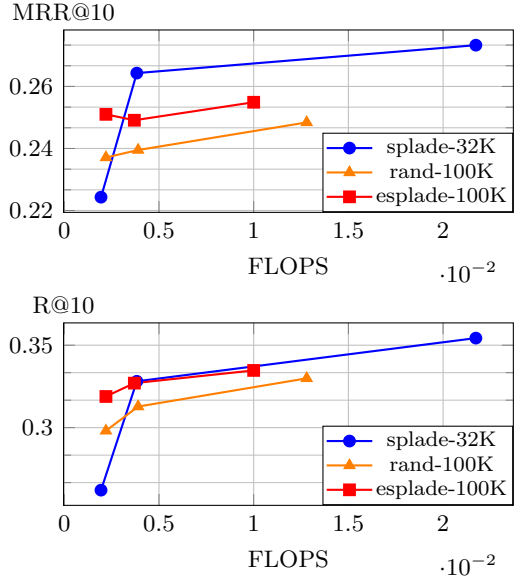

\subsection{Evaluation Result}

Figure \ref{tab:201} shows the evaluation result of finetuned models. Here we review the result by FLOPS level:

\subsubsection{Q, D Not Pruned}
In this setting, while the L0\_q is similar between models, the L0\_d of the 32K vocab model (splade-32K) is smaller than all 100K vocab models, while the FLOPS of the splade-32K is nearly two times higher than all 100K vocab models.
This means 100K vocab models can utilize larger vocabulary space, represent documents with a more diverse set of vocabulary, resulting inverted index with lower posting list length, and results in low FLOPS, despite the increased number of L0\_d. The Table \ref{tab:sample-ivt-stat} reflects this trend.
However, the retrieval effectiveness decreases when employing 100K models.
This means terms and scores from the splade-32K are better on retrieval effectiveness in this setting.

The rand-100K model shows decreased efficiency and slightly lower effectiveness than the esplade-100K model.
This can be interpreted as the randomly initialized MLM head can reach roughly similar performance to a pretrained EMLM head, but has limitations to learn semantically relevant sparse representation, hence requires more intersecting Q, D terms (higher FLOPS) to satisfy the ranking loss.
The esplade-100K model performs optimal FLOPS and best effectiveness among 100K models.
This means that the EMLM pretraining is essential for retrieval effectiveness and efficiency in terms of representing semantics of text more accurately, and can represent it more sparsely,  compared to its random counterpart.

\subsubsection{Q, D Pruned for FLOPS near 0.003x}
For this part,
where the model has a fixed number of activated Q, D terms at most by pruning, the FLOPS drops significantly across all models.
Both splade-32K and rand-100K models got dropped MRR and Recall performance. But not significantly as FLOPS drops, this means effective retrieval can be possible with high-scored terms (terms from top-k logits) only. This trend is similar to \cite{lassance2023static}.
The effectiveness of the esplade-100K model with qk=7 dk=20 option increases slightly compared to the unpruned one (Also, the MRR of qk=5 dk=10 model increased slightly compared to the qk=5 dk=20 model). This can be interpreted as terms beyond top-k terms (which have lower logit scores) can cause noise and could degrade model effectiveness.

This suggests that the full use (unpruned version) of representations trained from the N-pair metric learning loss objective is not always aware global structure of the embedding space \cite{oh2017deep},
hence does not always guarantee an optimal representation for retrieval on large-scale collections, which is affected by the distributions of entire documents.
The improvement can be interpretable as Q, D pruning alleviates such discrepancy of in-batch metric learning for retrieval.
However, obviously, when we prune more (qk=5 dk=20), both the MRR and Recall drop compared to the unpruned model.

\subsubsection{Q, D Pruned for FLOPS under bm25, near 0.002x}
For this part,
both the efficiency and effectiveness of splade-32K models drop significantly compared to their unpruned version.
~
A comparably smaller MRR and Recall drop occurred on rand-100K models compared to their unpruned version.
~
Which can be interpretable as dimensionality itself plays an important role for retrieval effectiveness, in terms of providing better discriminative power even if that dimension is initialized randomly.
This result is partly aligned with \cite{mackenzie2023exploring}, which higher vocab dimension leads to higher model effectiveness.\footnote{In the experimentation in \cite{mackenzie2023exploring}, the model of added-latent does not show a strong correlation of effectiveness to vocab dimension. In our experimentation (not shown here), we observed that added-latent vocabularies are not actively used for retrieval, as the posting list length of latent terms is smaller, and the score is smaller, compared to the regular vocabularies.
We expect the phenomenon to be caused by the prior weight of pretrained BERT, and the randomly initialized MLM FC area of added-latent vocabularies is less aligned, compared to the BERT weight and pretrained MLM FC area. This can lead the model to utilize added-latent terms less to satisfy the finetuning objective, relying more on the regular vocabularies. However, more experimentation and analyses are needed to clarify this clearly.}
~
This result can be interpretable as models with higher vocab dimensions are easy to maintain discrimination power at retrieval time by utilizing a more diverse set of vocabulary to represent Q and D.
At the same time, these models are easy to maintain efficiency since more vocab means the model can utilize the representational capacity of vocab indices more instead of vocab scores to obtain highly discriminative representations.
Such high vocabularic diversity of representation is observed to help achieve lower FLOPS, since it allows having more diverse terms to maintain representation powers in the pruning scenarios, leading to the smaller length of individual posting lists as depicted in the Table \ref{tab:sample-ivt-stat}.
For instance, in the "FLOPS under bm25" of Figure \ref{tab:201}, compared to the 32K model splade-32K, the L0\_d of 100K vocab models is increased by nearly twice (L0\_d=5 to L0\_d=9.91, L0\_d=9.96), while the FLOPS is increased by only 13.4\% (0.00194 to 0.0022).

The esplade-100K model maintains performance to its unpruned version, while being significantly efficient. This model has the same larger dimension of 100K vocabs as rand-100K models, but is pretrained.
According to the Figure \ref{tab:201}, in all cases, the esplade-100K models are effective than rand-100K models.
The esplade-100K is the same or more efficient for all cases.
This means EMLM pretraining, which involves BERT models to understand words in their contexts, helps to get a good initial weight for the finetuning stage, allowing the model to leverage larger dimensions better, including in the Q, D pruned setting.

\begin{table}
\small

\centering

  \begin{tabular}{l|ccc}
    \toprule
model and pruning strategy & mean & var & std\\
\midrule
splade-32K-d\_k\_full & 62.14 & 7196 & 84.83\\
esplade-100K-d\_k\_full & 19.5 & 1484 & 38.53\\
\midrule
splade-32K-d\_k20 & 41.91 & 3982 & 63.1\\
esplade-100K-d\_k20 & 13.03 & 959 & 30.97\\
\midrule
splade-32K-d\_k10 & 23.79 & 1497 & 38.69\\
esplade-100K-d\_k10 & 7.54 & 397 & 19.92\\

  \bottomrule
\end{tabular}
  \caption{\textbf{Stats of the individual posting list length for each model.} \small{The mean, variance, and standard deviation of the length of individual posting lists with different pruning strategies (d\_k is the remaining top k terms after pruning, where \_full means unpruned.) over 77,482 positive document titles of the validset.}}
  \label{tab:sample-ivt-stat}
\end{table}

\section{DISCUSSION}

\subsection{The Sole Effect of Vocab Size}

The SPLADE model shows better discriminative performance on both Recall and MRR in the default FLOPS setting (Q, D Not Pruned).
The ESPLADE model shows better discriminative performance in the low FLOPS settings (Q, D Pruned for FLOPS near 0.002x).
Low FLOPS means the traversal area of an inverted index is relatively small because the size of the inverted index and query representations are both reduced by pruning.
The experimental result shows that the ESPLADE model performs better when we constrain the traversal footprint of an inverted index to be smaller by pruning.
The SPLADE model's discriminative power is better when we traverse our index without constraining the traversal cost by pruning.

The structural difference between the SPLADE and the ESPLADE model is the vocab size.
One way to interpret the role of vocab size is by relating it to the contrastive finetuning for the ranking objective.
Conceptually, a larger vocab size can increase the model's representational capacity (i.e., a degree of the model's ability to approximate the target function) by providing additional output space (i.e., additional vocab and their scores) to construct more fine-grained decision boundaries for the training objectives.
This additional output space can help the model draw more flexible decision boundaries by leveraging the additional number of vocabularies, whereas a small output space draws such decision boundaries without the additional number of vocabularies, by only adjusting the MLM logit scores of existing vocabularies.
Hence, in splade training, we can consider one of the primary properties of a vocab size as the number of dots used to discriminatively represent Q and D in the global sparse embedding (or vector) space of the inverted index, apart from the vocabulary semantics in the first place.

There is recent work on increasing the vocabulary size of the models is helpful for NLP tasks \cite{kaya2024effect}, or emphasizes the importance of the domain-specific tokenization \cite{vaca2022named}.
However, the main focus of these works is studying vocab size and fine-granular computational expressivity of languages, which does not explain our performance improvement on the randomly initialized 100K-vocab model over the plain SPLADE model in the pruned setting.
We interpret this phenomenon is on a difference in task (broadly, NLP task vs IR task) and hence an inherent difference in their training losses (i.e., NLP Task Loss vs Ranking Loss with FLOPS Loss).
The outperformance of a randomly initialized 100K-vocab model proves that in LSR,
vocabs are used more than enhancing the accuracy of linguistic expressivity, but the size of the vocab itself performs solely a prominent role as a model's representational capacity, due to the training losses of the splade models.
This finding can enlighten different interpretations of the role of vocabulary for LSR.

\subsection{Ranking, FLOPS Losses Gradually Converts Wordpiece Vocab into Latent Vocab}

The experimental result implies vocabs of SPLADE models play beyond lexical expansion since the size of the vocabs is related to the representational capacity of the models after finetuning.
Current finetuning loss does not directly maintain the lexical or semantic meaning of sentences on their behalf, since it transforms sentence representation into representations for ranking where positive Q-D need to be similar; there are no direct positive D-D relations in the loss.
This can make vocabs partially lose their role in representing the lexical or semantics of sentences, since ranking representation needs to bridge different textual representations of Q and D into similar ones.
Moreover, the FLOPS regularization loss additionally damages the semantic fidelity of vocabulary by enforcing higher uniformity and sparsity for learning efficient sparse representations.

The ranking loss and FLOPS regularizer transform semantics of input Q, D text represented by pooled MLM logit vectors which is fed by pretrained BERT, or EMLM, or random MLM FC models, into ranking (Q-D similarity) aware and sparsity (uniformly distributed, retrieval efficient) aware representations in the sparse space defined by vocab dimension.
The initial MLM logit vectors are grounded from their literal representations, but this fine-tuning procedure gradually converts them to latent representations, anchoring different Q-D for ranking.
Hence, we can think of the vocabs of finetuned splade models are actually latent terms, and the vocab dimensions that define the number of latent term types available to represent Q, D, as a representational capacity of the model.
This interpretation may help to investigate that the SPLADE model has "wacky weight" on semantically less meaningful terms \cite{mackenzie2021wacky}. This interpretation can also help to understand the effectiveness of SPLADE models configured using controlled, semantically less meaningful vocab sets such as a set of stopwords, randomly sampled vocabs, low frequency terms, or latent (randomly initialized) terms as reported in \cite{mackenzie2023exploring}. In their report, finetuning splade of latent terms shows competitive performance to their counterparts, which have pretrained vocab vector and bias.
This can additionally support our interpretation.

In "Q, D Not Pruned" in Figure \ref{tab:201}, although there is some variance in L0\_d, values of L0\_q and L0\_d tend to converge into a similar volume in each, regardless of the models.
This is interesting because there are differences in vocab sizes and semantics.
Although further investigation is needed, we expect the combination of ranking and FLOPS loss affects this phenomenon, as the two losses treat vocab as a latent term instead of solely preserving their original semantics.

\subsection{Vocab Size, Pretraining, Retrieval Efficiency}

The result of Figure \ref{tab:201} shows that vocab size plays a crucial role in maintaining model performance when constraining the retrieval FLOPS by Q, D pruning.
On the other hand, the performance difference between esplade-100K and rand-100K shows the pretrained weights of the MLM head and BERT are another criterion for maintaining the model performance. This finding is aligned with reports from \cite{formal2024towards}.
There is literature discussing the relationships of MLM logit distributions and SPLADE finetuning.
For example, \cite{lassance2022efficiency} applied FLOPS regularization for MLM middle-training to obtain a more efficient model. 
\cite{nair2022learning} note an example of the distributional difference of MLM logits of pretrained models, which results in opposite finetuned performance \footnote{\cite{formal2024towards} reports the RoBERTa model \cite{liu2019roberta} does not finetune well. \cite{nair2022learning} reports the use of XLM-R model \cite{conneau2019unsupervised} shows low performance to mBERT \cite{devlin2019bert}. Understanding the distributional difference of these pretrained MLM and friendliness to ranking and FLOPS loss of splade finetuning seems to be beneficial. In the vocabulary, RoBERTa uses byte-level BPE, and XLM-R uses Sentence Piece model, where the BERT model uses Wordpiece.}.

As Table \ref{tab:sample-ivt-stat} shows smaller vocab model has a longer posting list; hence, Q, D pruning will result in larger skipping of potentially relevant document candidates and relevancy scores on those posting lists.
This results in sharp drops in both FLOPS and model effectiveness in the smaller vocab model in the Q, D pruned setting. 
This observation supports that vocab size in the splade models defines the structure of the inverted index and influences the retrieval efficiency.
However, finding optimal effective vocabulary size is a hyperparameter of learning, and seems inherently empirical, depending on the task, the dataset distribution.

In the "FLOPS under bm25" part in Figure \ref{tab:201},
the effectiveness of ESPLADE is higher compared to the random model, while the FLOPS of the two models are the same.
This means pretrained EMLM weights play an important role in achieving higher effectiveness within the same FLOPS.
However, as the previous study suggests \cite{lassance2023static}, we expect an increase in FLOPS to be generally inevitable to enhance the retrieval effectiveness, as FLOPS, a computational cost of retrieval, implies the intensity of Q, D interaction in the vector space to calculate the relevance.
In this context, high FLOPS essentially means effectively exploiting the representational capacity of the vector space retrieval model by giving more enriched Q, D representations.

\section{SUMMARY AND CONCLUSION}

In this paper, we finetuned the ESPLADE model with a 100K vocab size, and compared it to the model with a randomly initialized 100K vocab size and the plain SPLADE model with a 32K vocab size.
We conducted experiments on the evaluation set with strict query and document term pruning applied.
The experimental result shows that the larger vocab with randomly initialized weights shows increased retrieval effectiveness and retrieval cost compared to the plain SPLADE model which has a small, pretrained vocab.
The pretrained weights of the ESPLADE model further enhance retrieval effectiveness compared to the improved performance of the randomly initialized model, while maintaining the retrieval cost.
These results show that the vocab size is a key hyperparameter for pruning-based efficient retrieval in SPLADE models, and the pretraining on expanded vocabulary is essential to further improving the effectiveness.

Our observation demonstrates larger vocabulary size alone provides effective discriminative power of learning sparse representations by improving representational degrees of freedom, which leads to a performance gain by maintaining discriminative power when strict query and document pruning are applied.
We also revisited the importance of vocabulary content and corresponding pretraining, as it provides a better initial approximation of query and document representations for finetuning, which leads the model to maintain retrieval cost while being more effective.
Lastly, we discuss that the ranking and FLOPS regularization loss leverages the size and initial weights of vocabs to make representations for ranking, whether the vocab is pretrained or initialized randomly. Due to the two losses, we conjecture that Wordpiece vocab or randomly initialized vocab gradually converted to the ranking-centric latent vocab.

This suggests the role of vocabularies in learning sparse representations for ranking is to define the \textit{representational specification} of its neural representation for the retrieval engine, which processes the interaction, hence the communication of queries and documents.
This is a different view on the role of vocabulary in linguistics or natural language processing, where they use vocabulary to define the representational specification of lexical representations, a language for humans, and their communication.
We hope further investigation to be performed on this different role of vocabularies in LSR, which provides specification of queries and documents communications in the retrieval engine. Potential topics can include, but are not limited to, efficiency of training for larger dimensions, effective transferring of knowledge from a natural language corpus in pretraining, and so on.

\begin{table*}[b!]
\footnotesize
\begin{tabular}{lrrrrr}
\toprule
dataset & \# of q-title pairs & \# of unique titles & \# of queries & docs per query & queries per doc\\
\midrule
trainset-small & 182,784,463 & 79,154,465 & 21,394,466 & 8.543 & 2.104\\
\bottomrule
\end{tabular}
\caption{\textbf{Stats for trainset-small.}}
\label{tab:stats-set2}
\end{table*}
\begin{table*}[b!]
\centering

\setlength{\tabcolsep}{2.5pt}

\footnotesize
  \begin{tabular}{l|ccccc|cc|ccc|ccc}
    \toprule

model &  & vocab &         & dual & batch & \multicolumn{2}{c}{validset} & \multicolumn{3}{c}{validset} & \multicolumn{3}{c}{testset}\\

 &  type & size & Lj & loss &  size & L0\_q &  L0\_d &  MRR@10 &  R@10 &  R@100 &  MRR@10 &  R@10 &  R@100 \\
\midrule
bm25 & .  & . & . & . & . & . & . & 0.9427 & 0.7071 & 0.8112 & 0.7751 & 0.5066 & 0.6974 \\
splade-32K-ts & mlm & 32K & 3 & O & 7168 & 9.91 & 24.82 & 0.9446 & 0.739 & 0.8751 & 0.7363 & 0.4806 & 0.7263 \\
splade-32K & mlm & 32K & 3 & O & 7168 & 10.62 & 26.8 & 0.9461 & 0.7385 & 0.8699 & 0.7452 & 0.4859 & 0.7263 \\
\midrule
splade-32K-b3k & mlm & 32K & 3 & O & 3072 & 9.53 & 24.86 & 0.9345 & 0.7236 & 0.8684 & 0.695 & 0.4515 & 0.7082 \\
splade-32K-b3k-dlx & mlm & 32K & 3 & X & 3072 & 9.67 & 25.58 & 0.9332 & 0.7222 & 0.8685 & 0.6909 & 0.4487 & 0.7072 \\
rand-100K & rand & 100K & 5 & X & 3072 & 10.88 & 27.97 & 0.9267 & 0.7199 & 0.8709 & 0.6724 & 0.4373 & 0.7027 \\
esplade-100K & emlm & 100K & 5 & X & 3072 & 8.78 & 25.55 & 0.9312 & 0.721 & 0.8727 & 0.6854 & 0.4451 & 0.7043 \\
esplade-100K-lj3 & emlm & 100K & 3 & X & 3072 & 10.58 & 28.88 & 0.9394 & 0.7292 & 0.8749 & 0.7047 & 0.4584 & 0.7138 \\

  \bottomrule
\end{tabular}
\caption{\textbf{The validset and testset results of finetuned models.} \small{Models are trained on trainset. We put the splade-32K-ts in the table, which is trained on trainset-small. We retrain the baseline splade model with a smaller batch size for comparison. The validset result is based on the best R@10 checkpoint, and the testset result is based on the last step's checkpoint.}}
\label{tab:finetuning-result}
\end{table*}

\clearpage

\newpage

\newpage

\begin{acks}

We thank all members of the NAVER Search for their support.
We appreciate Young-In Song for his support of the team's research projects.
We thank Hyemin Lee for his support of the neural matching project over the year at the Image Ranking team.
No external funding was received in support of this work.
\end{acks}

\vskip 0.2in
\bibliography{sample-base}

\clearpage

\appendix

\small

\begin{table*}
\centering

\setlength{\tabcolsep}{3pt}
\footnotesize
  \begin{tabular}{l|ccccccc|ccccc}
    \toprule

model & \multicolumn{5}{c}{} & \multicolumn{1}{c}{topk} & \multicolumn{1}{c}{topk} & \multicolumn{5}{c}{validset}\\

 & q\_K & d\_K & L q & L d & L j & dual loss & mask KL & L0\_q & L0\_d & MRR@10 & R@10 & R@100 \\

\midrule

bm25 & . & . & . & . & . & . & . & . & . & 0.9427 & 0.7071 & 0.8112 \\
rand-100k-ts *1 *2 & 1000 & 2000 & . & . & 5 & X & X & 10.81 & 26.92 & 0.9285 & 0.7182 & 0.8663 \\
splade-32K-ts-0.1m-e1 & . & . & 5 & 0.2 & . & X & X & 9.67 & 31.03 & 0.947 & 0.7298 & 0.8648 \\
splade-32K-ts-0.1m-e2 & . & . & . & . & 5 & X & X & 8.05 & 20.31 & 0.9418 & 0.7262 & 0.8611 \\
splade-32K-ts-0.1m-e3 & 5 & 10 & . & . & . & X & X & 4.99 & 9.99 & 0.9277 & 0.6904 & 0.8343 \\
splade-32K-ts-0.1m-e4 & 10 & 20 & 5 & 0.5 & . & X & X & 6.83 & 14.76 & 0.9399 & 0.7222 & 0.861 \\
splade-32K-ts-0.1m-e5 & 100 & 200 & 5 & 2 & . & X & X & 6.92 & 13.16 & 0.9339 & 0.7168 & 0.8581 \\
splade-32K-ts-0.1m-e6 & 10 & 20 & . & . & 5 & X & X & 6.82 & 14.93 & 0.9381 & 0.7211 & 0.8599 \\
splade-32K-ts-0.1m-e7 & 5 & 10 & . & . & . & O & X & 5 & 10 & 0.9352 & 0.6987 & 0.8422 \\
splade-32K-ts-0.1m-e8 & 5 & 10 & . & . & . & O & O & 5 & 10 & 0.933 & 0.7004 & 0.841 \\
splade-32K-ts-0.1m-e9 & 10 & 20 & 14 & 2 & . & O & X & 5.15 & 10.37 & 0.9242 & 0.6919 & 0.8443 \\
splade-32K-ts-0.1m-e10 & 5 & 10 & 15 & 3 & . & O & X & 3.9 & 7.54 & 0.9089 & 0.6723 & 0.8333 \\
splade-32K-ts-0.1m-e11 & 10 & 20 & . & . & 15 & O & X & 4.98 & 11.51 & 0.9233 & 0.7004 & 0.848 \\
splade-32K-ts-0.1m-e12 & 10 & 20 & . & . & 18 & O & X & 4.72 & 10.86 & 0.9191 & 0.6933 & 0.8447 \\
splade-32K-ts-0.1m-e13 & 5 & 10 & . & . & 15 & O & X & 4.01 & 8.45 & 0.9132 & 0.6823 & 0.8371 \\
splade-32K-ts-0.1m-e14 & 10 & 20 & 5 & 1.5 & 5 & O & X & 4.58 & 8.99 & 0.911 & 0.6811 & 0.8399 \\
splade-32K-ts-0.1m-e15 & 5 & 10 & . & . & 18 & O & O & 3.87 & 8.2 & 0.9113 & 0.6762 & 0.835 \\
splade-32K-ts-0.1m & 500 & 1000 & . & . & 3 & O & X & 10.23 & 26.59 & 0.9451 & 0.7341 & 0.8645 \\
splade-32K-ts *1 & 500 & 1000 & . & . & 3 & O & X & 9.91 & 24.82 & 0.9446 & 0.739 & 0.8751 \\

  \bottomrule
\end{tabular}

\caption{\textbf{Validset result of different hyperparameter configurations.} \small{All models are trained on trainset-small. *1 denotes a model with 200K training steps, *2 denotes a model with 100K output vocab and a batch size of 3072. All models, except otherwise noted, follow these training configurations: The batch size of 7168, the learning rate of 1e-4, 16-bit (mixed) precision training, and the training step of 100K steps. The q\_K and d\_K are top-k masking from \cite{yang2021sparsifying}. The L q, L d is the FLOPS regularizer weight. The L j is the joint FLOPS regularizer weight. The dual loss and the mask KL are from \cite{yang2021sparsifying}. The validset result is based on the best R@10 checkpoint.}}
\label{tab:validset-pref}

\end{table*}

\section{HYPERPARAMETER AND PERFORMANCE}

\subsection{Validset Performance}

The Table \ref{tab:validset-pref} shows the Validset result of different hyperparameter configurations.
When comparing splade-32K-ts-0.1m-e3 and splade-32K-ts-0.1m-e7,
the top-k dual ranking loss \cite{yang2021sparsifying} appears to be effective.  
It seems that the supervision signal from the model without top-k masking helps guide the model that uses top-k.
When comparing the splade-32K-ts-0.1m-e9 and splade-32K-ts-0.1m-e11 pair, there doesn't appear to be a significant performance difference between using FLOPS loss and joint FLOPS loss in terms of the L0\_q, L0\_d.
This experimental result also shows that the model performance is proportional to the number of activation terms (L0\_q, L0\_d).
In the case of splade-32K-ts-0.1m-e11 and splade-32K-ts-0.1m-e12, performance is proportional to the number of activation terms.
Comparing splade-32K-ts-0.1m-e12, splade-32K-ts-0.1m-e13, and splade-32K-ts-0.1m-e10 the we can also observe that trend as well.

\clearpage

\begin{table*}

\centering

\setlength{\tabcolsep}{3pt}

\footnotesize
  \begin{tabular}{l|ccc|cc|ccc}
    \toprule

model & \multicolumn{3}{c}{} & \multicolumn{2}{c}{validset} & \multicolumn{3}{c}{Evaluation Set}\\

 & l\_method & q\_K & d\_K & L0\_q & L0\_d & FLOPS & MRR@10 & R@10 \\

\midrule

bm25 & . & . & . & . & . & 0.00277 & 0.203 & 0.2527 \\
splade-32K-ts-0.1m-e1 & org & . & . & 9.67 & 31.03 & 0.05002 (18x) & 0.2521 & 0.3209 \\
splade-32K-ts-0.1m-e1-postqk5-dk10 & org & . & . & 5 & 10 & 0.00623 (2.2x) & 0.2374 & 0.2898 \\
splade-32K-ts-0.1m-e2 & joint & . & . & 8.05 & 20.31 & 0.01797 (6.4x) & 0.2392 & 0.3035 \\
splade-32K-ts-0.1m-e2-postqk5-dk10 & joint & . & . & 5 & 10 & 0.00410 (1.4x) & 0.2292 & 0.2831 \\
splade-32K-ts-0.1m-e4 & org & 10 & 20 & 6.83 & 14.76 & 0.01815 (6.5x) & 0.2369 & 0.3055 \\
splade-32K-ts-0.1m-e12 & joint & 10 & 20 & 4.72 & 10.86 & 0.00441 (1.5x) & 0.2104 & 0.271 \\
splade-32K-ts-0.1m-e8 & none & 5 & 10 & 5 & 10 & 0.01634 (5.8x) & 0.1977 & 0.2553 \\
splade-32K-ts-0.1m-e15 & joint & 5 & 10 & 3.87 & 8.2 & 0.00204 (0.7x) & 0.1776 & 0.2339 \\
splade-32K-ts-0.1m-e5 & org & 100 & 200 & 6.92 & 13.16 & 0.01487 (5.3x) & 0.2377 & 0.3003 \\
splade-32K-ts-0.1m-e6 & joint & 10 & 20 & 6.82 & 14.93 & 0.01085 (3.9x) & 0.2259 & 0.2927 \\
splade-32K-ts-0.1m & joint & 500 & 1000 & 10.23 & 26.59 & 0.03110 (11.2x) & 0.2449 & 0.3136 \\
splade-32K-ts-0.1m\_postqk5-dk10 & joint & 500 & 1000 & 5 & 10 & 0.00460 (1.6x) & 0.2324 & 0.2823 \\
rand-100k-ts & joint & 500 & 1000 & 10.81 & 26.92 & 0.01777 (6.4x) & 0.2153 & 0.2838 \\
rand-100k-ts\_postqk5-dk10 & joint & 500 & 1000 & 5 & 10 & 0.00272 (0.98x) & 0.2 & 0.2458 \\

  \bottomrule
\end{tabular}
\caption{\textbf{Result on the Evaluation Set.} \small{l\_method is the regularization loss method, org is the original FLOPS loss used in the splade paper \cite{formal2021splade}, and joint is the joint FLOPS loss presented in \cite{dudek2023learning}. The -post suffix on the model name represents the static pruning option applied to queries and documents, by remaining only the given top-k query and document terms. Here, the L0\_q and L0\_d are measured against the validset. The validset result is based on the best R@10 checkpoint.}}
\label{tab:finetuned-model-eval-result}
\end{table*}

\subsection{Evaluation Set Performance}

The Table \ref{tab:finetuned-model-eval-result} shows the Evaluation Set result of models.
As the performance of model splade-32K-ts-0.1m-e15 shows, strict top-k masking during training results in very low activated terms, which hurts effectiveness a lot. This indicates that the model can not leverage enough vocabulary to represent queries and documents during training.
~
When a regularizer does not exist (in the case of splade-32K-ts-0.1m-e8), the effectiveness is increased compared to splade-32K-ts-0.1m-e15, but the efficiency decreases a lot.
This verifies that the utilization of vocabulary space by FLOPS regularizer on finetuning is crucial for constructing efficient representation, even in the constrained top-k masked training settings.
~
~
~
Compared to the splade-32K-ts-0.1m-e2-postqk5-dk10 model, the splade-32K-ts-0.1m\_postqk5-dk10 model increases FLOPS slightly and with similar model performance. This indicates that large top-k masking values does not significantly affect performance of the model. This indicates that lower MLM logits are not much used for training \footnote{Currently, the size of MLM logits causes training bottlenecks when the vocabulary is large. Methods for skipping the saving of lower-ranked MLM logits in the GPU memory can be helpful for training models with larger vocabularies.}.

\end{document}